\documentclass[pt12]{iopart}
\pdfoutput=1
\usepackage{iopams}
\usepackage[usenames,dvipsnames]{color}

\usepackage{graphicx}

\begin{document}

 \title[Thermodynamic limits to information harvesting by sensory systems]{Thermodynamic limits to information harvesting by sensory systems} 
 \author{Stefano Bo$^1$, Marco Del Giudice$^1$,
Antonio Celani$^{2}$}
\address{$^1$ Physics Department, University of Turin and INFN, via P. Giuria 1, I-10125 Torino, Italy}
\address{$^2$ The Abdus Salam International Centre for Theoretical Physics (ICTP), Strada Costiera 11, I-34014 - Trieste, Italy}
\ead{stefano.bo@ircc.it}
 

\begin{abstract}
In view of the relation between information and thermodynamics we investigate 
 how much information about an external protocol can be stored
in the memory of a stochastic measurement device given an energy budget.
We consider a layered device with a memory component
storing information about the external environment 
by monitoring the history of a sensory part
coupled to the environment. 
 We derive an integral fluctuation theorem for the entropy production and a measure
 of the information accumulated in the memory device. Its most immediate consequence is that the 
 amount of information is bounded by the average thermodynamic entropy produced by the process.
At equilibrium no entropy is produced and therefore the memory device does
not add any information about the environment to the sensory component.
Consequently, if the system operates at equilibrium the addition of a memory component
is superfluous. 
Such device can be used to model the sensing process of a cell
measuring the external concentration of a chemical compound 
and encoding the measurement in the amount of phosphorylated cytoplasmic proteins.
\end{abstract}
\date{\today}

\maketitle

\paragraph{Introduction.---}

Information and thermodynamics are deeply intertwined. 
Knowledge of the microscopic state of a system allows to extract energy beyond the
bounds set for systems for which no information is available.
Information can therefore provide energy (see Ref. \cite{maruyama} for a review).
Conversely, managing information has a thermodynamic cost in terms of entropy production
 \cite{landauer61}. 
 
Recently, the study of systems where fluctuations play a prominent role has 
provided an excellent ground for testing and developing these fundamental considerations.
When such systems allow a theoretical description in terms of stochastic processes, 
it is possible to derive exact relations connecting information and energetics (See Refs. 
 \cite{seifert_review,sekimoto,ritort08,esposito10,sagawa_ueda09,kawai07,sagawa_ueda10,sagawa_ueda12,sagawa_ueda13,Ito_Sagawa,Pekkola,
horowitz_sag,Horowitz_Vaik,horowitz_esposito,granger11,mandal12,still12,tusch14,allahverdyan09,
barato14prl,barato13jstatphys,hartich14} for a representative, yet necessarily incomplete list of contributions). 
The increasing experimental ability to measure and control the
evolution of small systems has led to remarkable verifications of these newly derived results
\cite{berut12,toyabe10,koski14}. 

One of the most natural fields 
of application for these ideas is cell biology. Indeed, cells, either as single organisms or in a multicellular assembly, need to 
acquire, exchange and process information to be able to survive and prosper, and these processes are permeated with noise (see Ref. \cite{bialek} for a general discussion and \cite{bowsher14} for a recent review
on information studies in biology).
Information transfer consumes energy and 
a quantitative comparison between
the costs (energy consumption) and the benefits (information acquired) might shed light on the tradeoffs that 
 have shaped the cellular architecture. In this framework, biological processes such as
gene regulation \cite{tkacik09}, cellular
switches \cite{qian05,tu08}, kinetic proofreading \cite{murugan12,sartori_pigolotti} 
have been considered in detail.
In addition, the problem of sensing in single-celled organisms has attracted particular attention \cite{berg_purcell,bialek05,govern12,endres09,hu10}. The efforts to relate the energy
spent by a cell to measure some features of the external environment to the precision of the measurement, or to the 
amount of information acquired, have produced several interesting results \cite{mehta,lan12,sartori14,tenwolde,skoge12,becker13,barato13pre,barato14,lang14}.
The present contribution carries on along this line of reasoning and aims at providing a general relationship between energy expenditure
and harvested information in sensory systems. 

\paragraph{Formulation of the problem.---}
We consider a measurement device composed of a sensory layer (labeled as $y$)  
which evolves according to a Markov process whose transition probabilities 
depend on the state of the external environment (labeled as $\lambda$) which may depend on time,
and a memory layer, attached to the sensory part, that reads, processes and stores its output $y$ in a state $x$. 
The full system $(x,y)$ is a Markov chain (see Figure~1a). 
\begin{figure}
\includegraphics[width=\columnwidth]{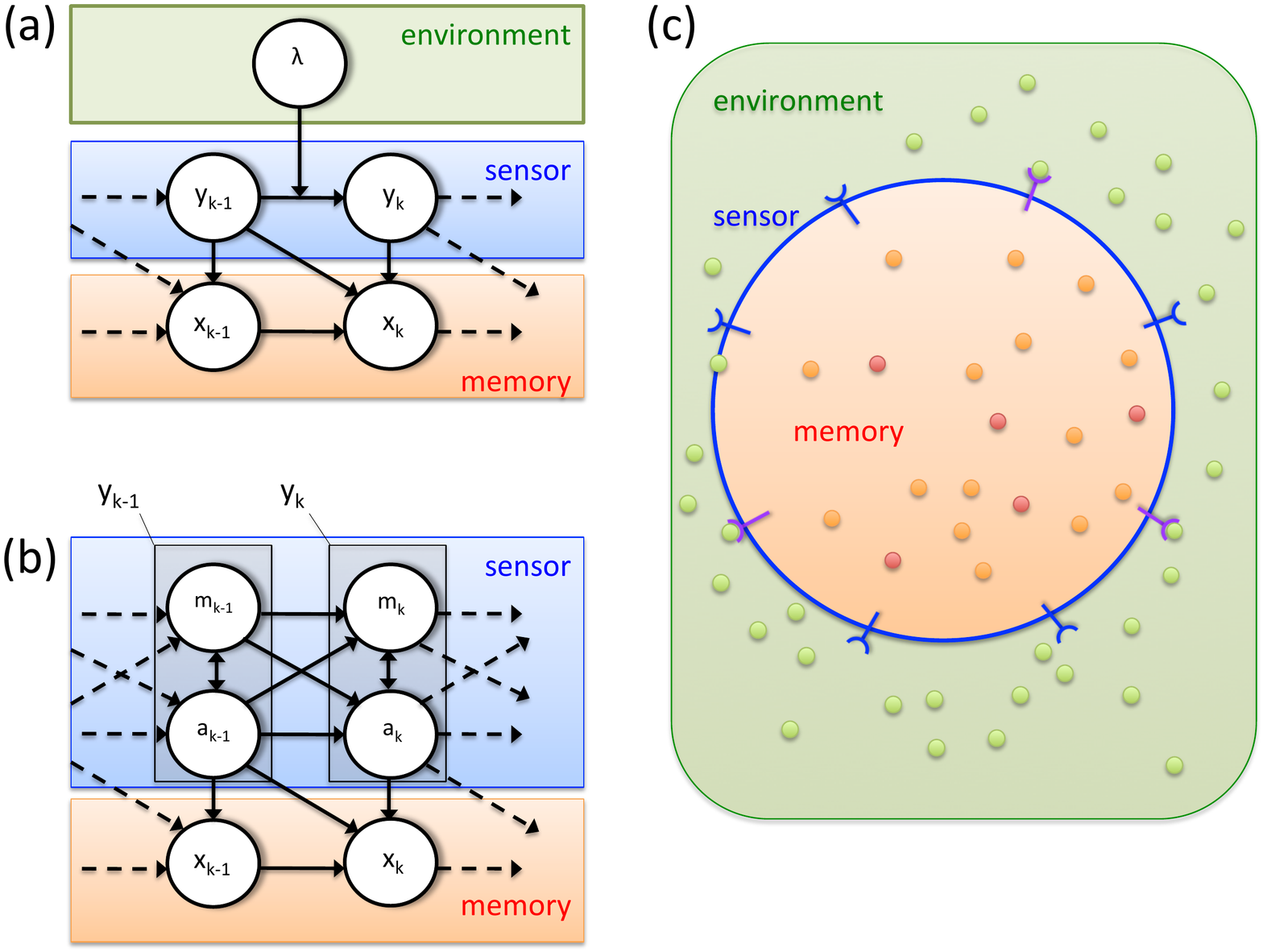}\label{fig:drawing}
\fl \caption{Stochastic measurement devices. (a) The state of the environment $\lambda$ influences the transition rates of the sensory layer $y$, which are however unaffected by the states of the memory $x$. The transitions in the memory $x$ depend on the previous and current state of the sensory part. (b) The sensory layer may have a complex structure  with internal feedbacks, as long as 
the memory does not affect its dynamics. We have omitted the environment for the simplicity of the drawing. This specific example depicts a minimal model for the chemosensory system in {\it E. coli} \cite{Berg}. The sensory layer is composed by a pair  $y=(m,a)$ where $m$ is the methylation state of the receptor cluster and $a$ is its activity (phosphorylation of the kinase CheA). The states of the memory are the number of phosphorylated second messenger proteins CheY. (c) A cell viewed as a measurement device. Bound receptors are shown in violet, unbound ones in blue, activated proteins in red, inactivated proteins in orange. Each receptor can activate or deactivate a single protein with a rate that depends on whether it is bound or unbound.}
\end{figure}

The dynamics of the sensory layer is not affected by the state of the memory device, i.e. no direct feedback of the output of the 
computation on the sensory system is considered. This is a crucial assumption, whose implications and limitations are addressed in the discussion at the end of the paper. (Notice, however, that this requirement is not as restrictive as it may seem at first glance. For instance, it perfectly accommodates for all cases where
there is feedback within the sensory layer, e. g. in the {\it E. coli} chemosensing system. See Figure 1b  and Ref.~\cite{Berg}).

A simple, biologically relevant example is when the sensory system is a set of receptors that bind an external ligand and the 
memory is a pool of cytoplasmic proteins that can be activated, e.g. phosphorylated, by the receptor (see Figure 1c and Ref.~\cite{mehta}) . In this case $\lambda$ is the external concentration, $y=0,\ldots,R$ is the number of bound receptors and $x=0,\ldots,N$ is the number of activated proteins. This specific example will be discussed in detail below.

In this framework, the fundamental question that we address here is: given an energy budget, how much information about the external environment $\lambda$ can be stored in the {\it current} state $x$ of the memory~?

\paragraph{The history of detections.---}
It is important to note that the external environment affects the 
whole time evolution of the sensory layer.
As a consequence, it is the sequence of detections made by the sensory part that encodes knowledge 
about the environment, rather than simply its current state. 
To illustrate the concept that knowledge of the past history of detections is a fundamental ingredient for the process of cellular decision-making we offer two examples.

Cells constantly measure concentration levels of external ligands. A statistically efficient approach is 
to perform a maximum likelihood estimation 
making use of the time series of receptor occupancy \cite{endres09}. The best estimate of the binding rate, which is proportional to ligand concentration, is then obtained as the ratio of the number of 
 binding events to the time spent in the unbound state. This estimate clearly depends on the specific sequence of occupancy and converges very slowly, as the inverse square root of time, to its expectation value. Therefore the specific history of detection matters a lot in the estimate, especially when decisions have to be taken quickly.

The dynamical character of information harvesting is even more apparent when the cell has to decide on the fly between two choices, e.g. whether the environment is either $\lambda$ or $\lambda'$, given maximum allowed values for the error. 
When the decision is difficult, this task is optimally accomplished by the Wald ratio test, which requires the evaluation of the log-likelihood ratio of the sequence of receptor states (see Ref.~\cite{siggia13}). 

Upon having agreed that the state of the memory device has not only to record the current state of the sensory layer
but also encode knowledge about its history, the question is then how to identify the proper measure for this information about the past.

\paragraph{The amount of harvested information.---}

For the sake of generality, hereafter we discuss the case when the stochastic system is a discrete-time Markov chain.
The results extend obviously to continuous-time Markov processes and diffusive systems.  
The amount of information present in the current state of the system $(x_n,y_n)$ about the past detections of the sensory layer
$\{y_0,\ldots,y_{n-1}\}$ can be written in terms of the pointwise mutual information 
$ i(\{x_n,y_n\}:\{y_0,\ldots,y_{n-1}\}) \equiv \log  \frac{p(x_n,y_n|y_0,\ldots,y_{n-1})}{p(x_n,y_n)}$ (see e.g. \cite{cover}).
This quantity can be split in the sum $i(x_n:\{y_0,\ldots,y_{n-1}\}|y_n)  + i(y_n:\{y_0,\ldots,y_{n-1}\})$,
where the first term is the information about the past detection history contained in the output, given the present 
state of the sensory layer, and the second one is the information that the sensory layer alone has about its past.
We will focus on the information 
\begin{equation}\label{eq:info}
 {\cal I}_n = i(x_n:\{y_0,\ldots,y_{n-1}\}|y_n) 
=i(x_n:\{y_0,\ldots,y_n\})-i(x_n:y_n)
\end{equation}
which measures the amount of memory that the $x_n$ state alone has about the past history of the sensory layer,
excluding the information it has about $y_n$. 
The information ${\cal I}_n$ is a measure of the additional information content about the external 
environment warranted by the presence of a memory layer. 

As a side remark, let us note that some recent studies (see Refs. \cite{Ito_Sagawa,horowitz_sag,hartich14})
have addressed the relation between information
and thermodynamics in sensing by considering the transfer entropy defined in \cite{schreiber}. 
Our focus is on the information 
on the history of the sensory layer added at a given time by the presence of a memory. 
We remark that this differs from the transfer entropy which is often used to account for the information 
transfer in systems with a 
feedback.
More in detail, the authors of \cite{Ito_Sagawa} considered a combination of initial mutual information, final mutual information
and transfer entropy which again is different from the one we have defined and that at variance with ours is not 
positive on average.

The average value of ${\cal I}_n$
can be expressed (see the proofs section) in terms of the Kullback-Leibler distance between the probability
distribution of the memory states conditioned
on the whole history of the sensory layer and the one conditioned on the current
state only:
\begin{eqnarray}\label{eq:info_kl}
 \left\langle {\cal I}_n\right \rangle 
 = \sum_{y_{0}\ldots y_{n}} p(y_0,\ldots,y_{n}) D_{KL} \left( p(x_n|y_0,\ldots,y_{n}) \Bigr|\Bigr| p(x_n|y_n) \right)\;.
\end{eqnarray}
Indeed, when ${\cal I}_n=0$, there is no point in having a memory unit 
$x_n$ as the information about the past history of detections $\{y_0,\ldots,y_n\}$
is the same as the one contained in the current state of the sensory layer $y_n$.
By definition, ${\cal I}_n$ vanishes at $n=0$, it has a positive average, and is
bounded by the Shannon entropy of the the memory device 
$0 \leq\langle {\cal I}_n\rangle\leq H(x_n|y_n) \leq H(x_n)$, 
which is ultimately bounded by the logarithm of the number of states that the memory layer can take.
While the choice of this specific quantity may appear reasonable yet arbitrary at this point, it will become more than justified below,
where we will  prove that ${\cal I}_n$ appears together with entropy production in an
integral fluctuation theorem. 
%
%
%
%

\paragraph{An information-dissipation identity.---}
The entropy production of the measurement device after $n$ steps is \cite{schnakenberg,hill,barato14,seifert_review,ritort08}
\begin{equation}\label{eq:entropy_full}
{\cal S}_n  =-\log \frac{p(x_n,y_n)}{p(x_0,y_0)} + \sum_{k=0}^{n-1} \log \frac{p(x_{k+1},y_{k+1}| x_k,y_k)}{p(x_{k},y_{k}| x_{k+1},y_{k+1})}
\end{equation}
Above, $p(x_k,y_k)$ is the probability that the system is in state $(x_k,y_k)$ at time $k$ and $p(x_{k+1},y_{k+1}|x_{k},y_{k} )$ is the transition probability
from state $x_{k},y_{k} $ to state $x_{k+1},y_{k+1}$. (The dependence on the external environment $\lambda$ is implied). The entropy production of the sensory layer alone is
\begin{equation}\label{eq:entropy_sens}
{\cal S}^{Y}_n=\log p(y_0)-\log p(y_n) + \sum_{k=0}^{n-1} \log \frac{p(y_{k+1}|y_k)}{p(y_{k}| y_{k+1})}\;.
\end{equation}
Both entropies obey the known fluctuation theorems
$\langle e^{-{\cal S}_n}\rangle=1$ and   $\langle e^{-{\cal S}^{Y}_n}\rangle=1$, (see e.g.  Refs.~\cite{seifert_review,ritort08})
where $\langle\; \rangle$ denotes the average over all possible sequences.

Our main result that links entropy production to information is the identity
\begin{equation}\label{eq:IFT}
 \left\langle e^{-{\cal S}_n + {\cal S}^{Y}_n + {\cal I}_n } \right\rangle =1\;. 
\end{equation}
By Jensen's inequality, it implies the  bound
\begin{equation}\label{eq:ineq}
\left\langle {\cal S}_n \right\rangle -\left\langle {\cal S}^{Y}_n \right\rangle  \ge  \left\langle {\cal I}_n \right\rangle \ge 0\;.
\end{equation}
Therefore, the information that the memory device can store about the history of detections
is bounded by the average entropy production of the full process minus the average entropy
produced by the sensory layer alone.
The proofs are given in the proofs section.
%
\paragraph{Sensory devices at equilibrium do not harvest information.---}
A first relevant consequence of this inequality is that the system must be out of equilibrium in order to store information.
Indeed, if the system is at equilibrium it does not produce entropy and a straightforward consequence of (\ref{eq:ineq}) is that $\langle {\cal I}_n \rangle = 0$.  Similar results involving the impossibility for cells to measure at equilibrium for specific systems
have already been found for other definitions of information or accuracy 
(see e.g. \cite{mehta,lang14,tenwolde,barato14,horowitz_esposito}).

 The statement that no information can be added to the sensory layer at equilibrium can be understood more intuitively as well.
 Since we are considering a system with no feedback, the future evolution of the sensory layer is independent
 of the current state of the memory device. Hence, 
 $I\left(x_k:\{y_{k},\ldots y_n\}\right)=I(x_k:y_k)$ 
 (see the proofs section).
 If the system is at equilibrium, its evolution is time-reversible and therefore
 having no information on the future corresponds to having no information about the past.
 As a consequence, at equilibrium 
 $ I\left(x_k:\{y_1, \ldots, y_k\}\right)=I\left(x_k:y_{k} \right)  $
and therefore
$  {\cal I}_k=0 $ for all $k$, making the memory device totally worthless. 

An even stronger result can be obtained
 for systems that, in addition to having no feedback from memory, also do not display transitions
 that change simultaneously the states of both variables. These are dubbed
 bipartite systems (see Refs. 
 \cite{barato13pre,barato13jstatphys,barato14,hartich14,diana,horowitz_esposito}).
 In this case we prove (see the proofs section)  that, at equilibrium, the memory part does not even possess information 
 about the current state of the sensory device, i.e. the $X$ process is completely independent of the $Y$ process
 $ I\left(x_k:\{y_1, \ldots, y_n\}\right)=I\left(x_k:y_{k} \right)=0$.
 
 \paragraph{Learning and dissipation rates.---}
 We now revert to the general out-of-equilibrium case to discuss further implications of the information-dissipation inequality.
Since (\ref{eq:ineq}) applies to all times $n$  one also has the bounds
\begin{equation}
 \sup_{n\ge 1} \frac{\left\langle {\cal S}_n -{\cal S}^{Y}_n \right\rangle}{n} \ge 
\sup_{n \ge 1} \frac{\left\langle {\cal I}_n\right \rangle}{n} \ge \left\langle {\cal I}_1\right \rangle \;.
\end{equation}
This result becomes particularly informative for a system at steady state
since in this case the l.h.s. equals the average entropy production rate ${\sigma}=\langle {\cal S}_1-{\cal S}_1^Y\rangle$
and 
\begin{equation}\label{eq:learn}
\sigma \ge  \left\langle {\cal I}_1\right \rangle = \sum_{x,y,x',y'} p_{ss}(x,y;x',y') \log \frac{p_{ss}(x'|y,y')}{p_{ss}(x'|y')}
\end{equation}
recovering the bounds on learning rate discussed in Refs. \cite{horowitz_esposito,barato14,hartich14} for bipartite systems.
At the steady state, we show that $\langle {\cal I}_n \rangle$ 
is an increasing function of time (see the proofs section).
Information is not forever accrued, though, as it is bounded from above by the entropy of the memory device
 $\langle {\cal I}_n \rangle \le I(x_n:\{y_0,\ldots,y_n\}) \le H(x_n)$. Therefore, at the steady state energy dissipation increases linearly in time while the total harvested information reaches a finite value because of the limited memory that can be stored in the device.

\begin{figure}[!h]

\includegraphics[
width=\columnwidth
]{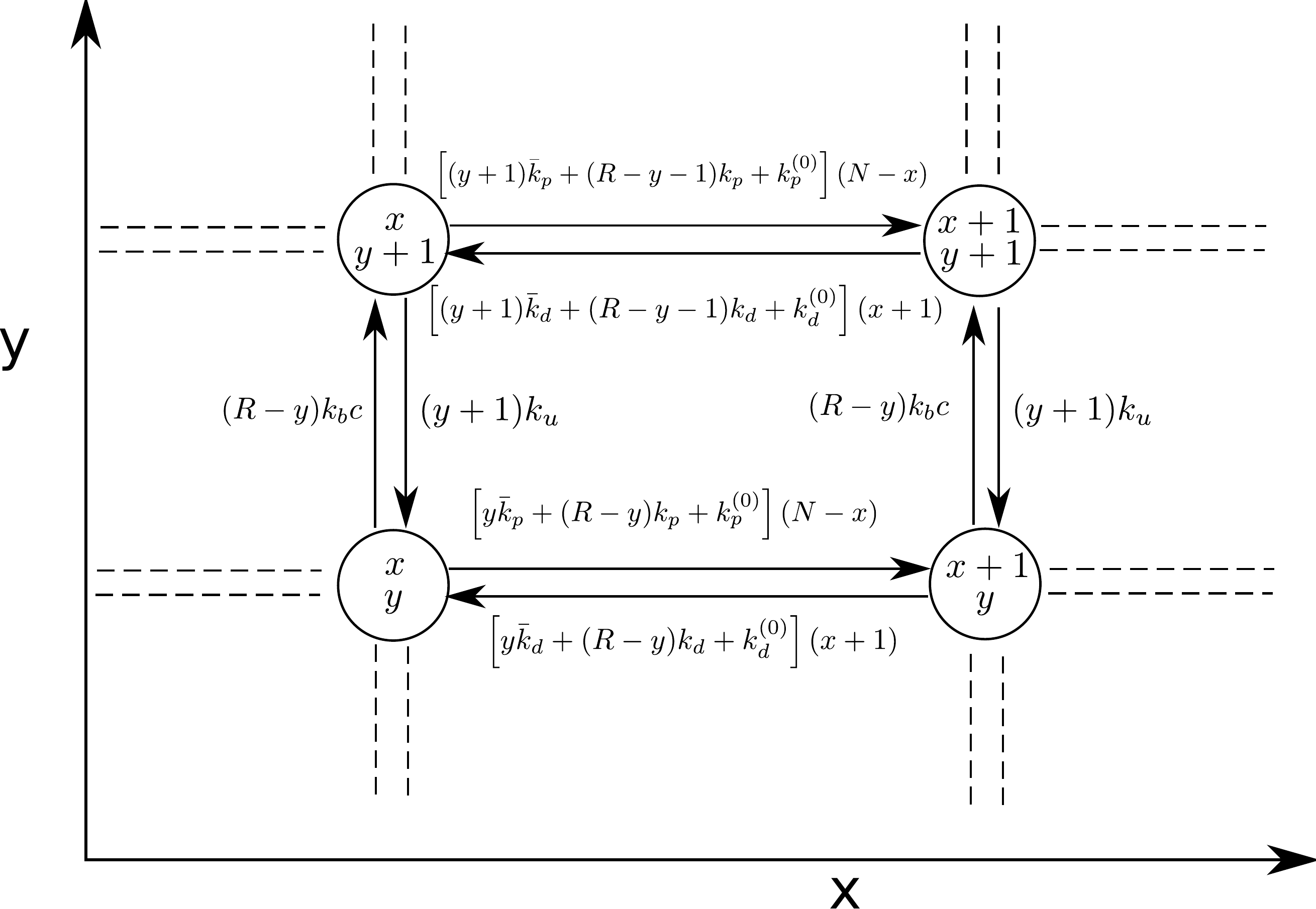}

 \caption{Schematic view of the chemical reaction network described by the rates of Eq.~(\ref{eq:rate1}) and Eq.~(\ref{eq:rate2}).}

\label{fig:me}

\end{figure}

\paragraph{Biochemical sensory devices.---}
In order to illustrate our results let us focus on
a biologically motivated example (see Figure 1c). 
Expanding on the model discussed in Ref.~\cite{mehta} we consider a  biochemical sensory
system made of a collection of $R$ independent receptors that bind a ligand at concentration $c$, 
 and a
computation device made of a finite pool of $N$ cytoplasmic proteins that can (de-)activate (e.g. (de-)phosphorylate) either spontaneously 
or through the action of the receptor (see Fig.~\ref{fig:me}).
This system is described by a chemical master equation, i.e. a continuous time Markov process (see Ref. \cite{gillespie_ME})
with transition rates 
\begin{equation}\label{eq:rate1}
 K(x',y' \to x,y) = \chi(y' \to y) \delta_{x,x'} + \xi(x' \to x|y) \delta_{y,y'}
 \end{equation} 
 where 
\begin{eqnarray}\label{eq:rate2}
\chi(y'\to y) =  \delta_{y,y'+1} k_b c  (R - y') + \delta_{y,y'-1}  k_u y' \\\nonumber
\xi(x' \to x|y) =\delta_{x,x'+1} \hat{k}_p(y)  (N - x') + \delta_{x,x'-1} \hat{k}_d(y) x' \; .
\end{eqnarray}

Since the chemical reactions change either the number of bound receptors or that of phosphorylated proteins 
this system is bipartite.
The activation/deactivation rates depend on the receptor state as
$\hat{k}_{p,d}(y)= \overline{k}_{p,d} y+k_{p,d} (R-y) + k_{p,d}^{(0)}$. Here, $\overline{k}_{p,d}$ 
is the (de-)phosphorylation rate by a bound receptor, $k_{p,d}$ by an unbound receptor, and $k_{p,d}^{(0)}$ the rate for spontaneous (de-)activation.
The master equation for this system then reads
\begin{eqnarray}\label{eq:master_eq}
 \frac{dp(y,x)}{dt} &=&(y+1)k_u p(y+1,x)+(R-y+1)ck_bp(y-1,x)\\\nonumber
&+&(x+1)\hat{k}_d(y)p(y,x+1)+(N-x+1)\hat{k}_p(y)p(y,x-1)
 \\\nonumber
 &-&\left[x \hat{k}_d(y) +(N-x)\hat{k}_p(y)+yk_u+(R-y)ck_b\right]p(y,x)\,.
\end{eqnarray}
If the rates obey $\frac{\overline{k}_p}{\overline{k}_d} = \frac{k_p}{k_d} =\frac{k_p^{(0)}}{k_d^{(0)}} =K$ and the external ligand concentration
does not change, the system relaxes to an equilibrium steady state and, as shown in the previous sections, measurement is impossible.
For any other choice of the rates the system is out of equilibrium and therefore it produces entropy and it is able to acquire information about the
environment.
If the concentration of the external ligand does not change, the system relaxes exponentially to
a steady state. For the sake of clarity we
consider such stationary settings.

\begin{figure}[!h]
\centering

\includegraphics[
width=0.8\columnwidth
]{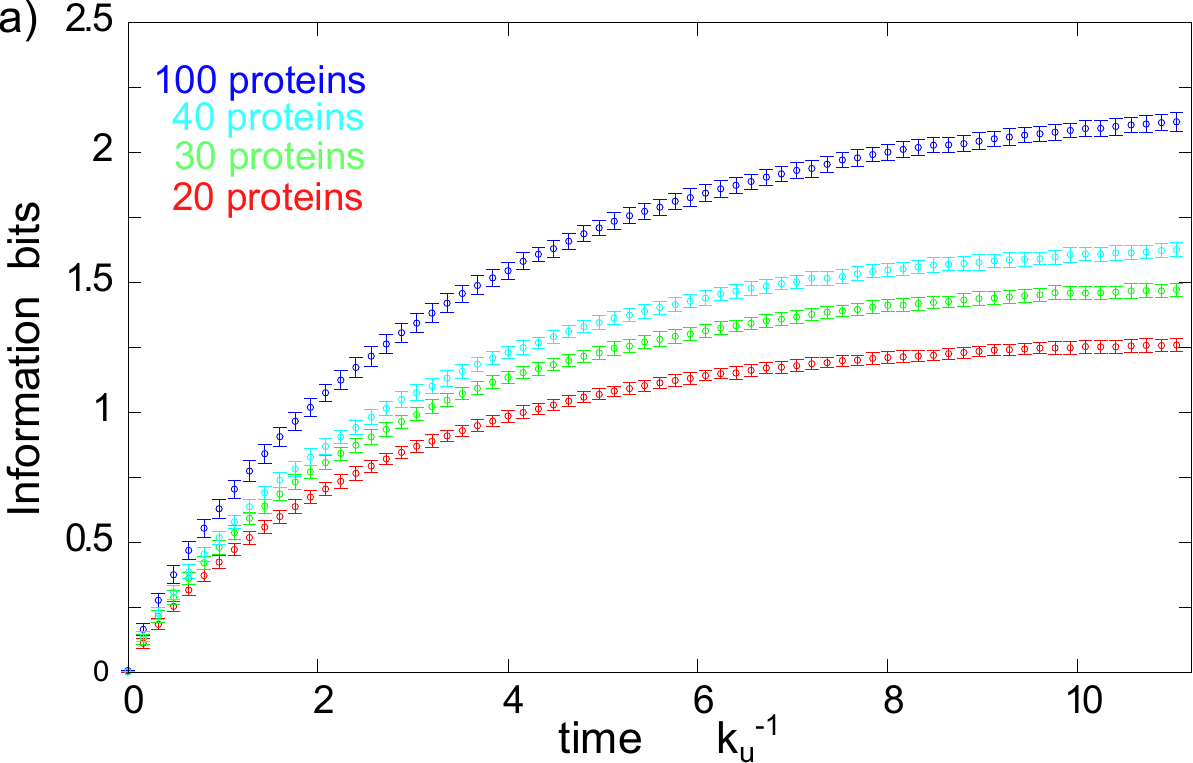}
\includegraphics[
width=0.8\columnwidth
]{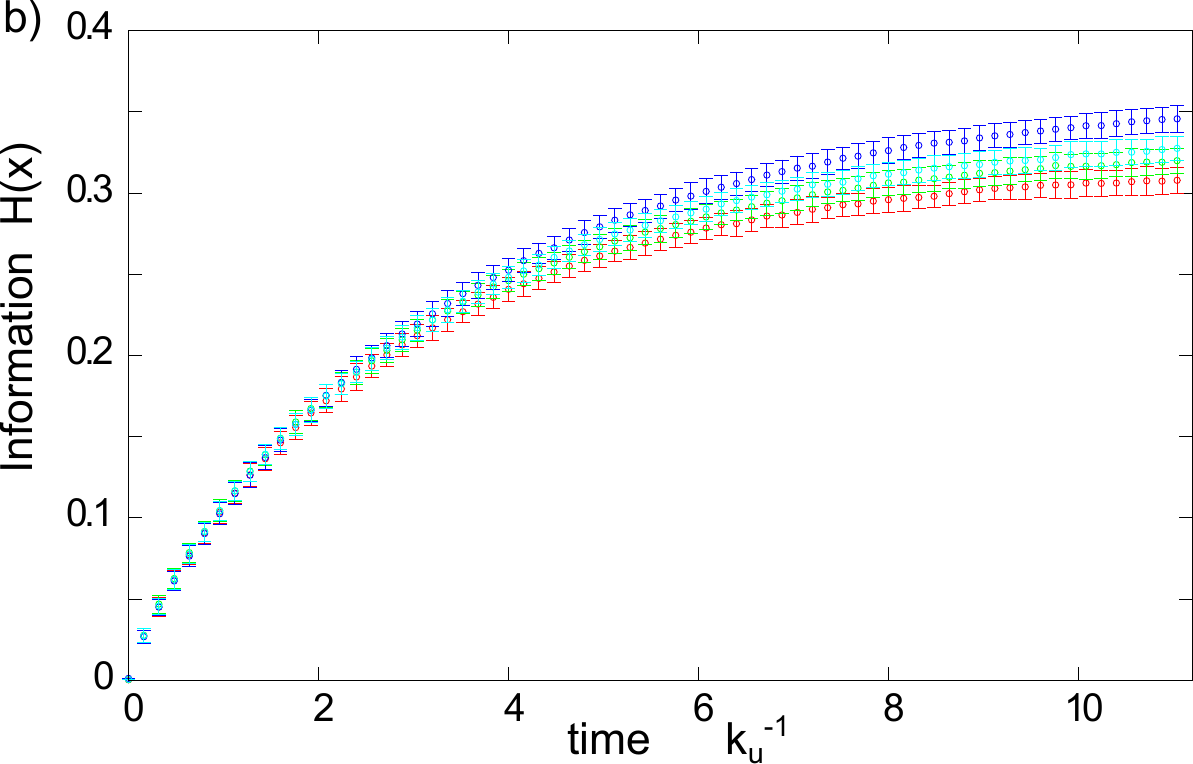}

 \caption{
(a) Harvested information at the steady state (in bits) as a function of time (in units of $k_u^{-1}$) for different numbers of proteins $N$ and $1$ receptor.
(b) Same plot rescaled by the Shannon entropy of the protein distribution $H(X)$. Information is now in units of $H(X)$.
The parameter set corresponds to the erasure procedure discussed in the text:
$k_u=k_b=0.2$,  $\bar{k}_p=100k_u$, $\bar{k}_d=0.05k_u$, $k_p=0.005k_u$, $k_d=0.15k_u$, $k_{p}^{(0)}=k_{d}^{(0)}=0$ with an 
external ligand concentration  $c=0.2k_u/k_b$.}

\label{fig:numerics_many}

\end{figure}
 {  
Entropy production can be related to the thermodynamic forces at play.
For the system we are considering the thermodynamic force keeping it out of equilibrium is provided by
 the  ATP hydrolysis which drives protein phosphorylation
  only when the receptor is bound (see e.g. Ref.~\cite{barato14}).
It is important to point out that, generally speaking, Eq.~(\ref{eq:master_eq}) implies that there are several chemical reactions that connect a given pair of initial  and final states -- i.e. given initial and final numbers of bound receptors and phosphorylated proteins. In this case, the correct definition of entropy production requires to separately account for all the available transition channels (reactions), as clearly discussed in \cite{esposito10}. However, below we limit ourselves to the case 
of a single receptor
($R=1$) and no spontaneous  (de-)phosphorylation ($k_{p}^{(0)}=k_{d}^{(0)}=0$): in this case each transition corresponds to a single chemical reaction. As a result,
the entropy production of the Markov system coincides with the chemical entropy production. 

Further assuming that the receptor reaches equilibrium, so that at the steady state it does not produce entropy (${\cal S}_1^{Y}=0$), the entropy production rate of the full system is
\cite{seifert_review,esposito10}:
\begin{equation}\label{eq:bc_entro}
\sigma ={\cal S}_1 = \sum_{x,y,x',y'} p_{ss}(x,y)K(x,y \to x',y') \log \frac{K(x,y \to x',y')}{K(x',y' \to x,y)}
\end{equation}
}
where $p_{ss}$ denotes the steady state probability.

For the sake of clarity let us start by considering the case of one receptor and one protein.
Thermodynamic consistence then dictates 
{ 
\begin{equation}
 \log{\frac{\overline{k}_pk_d}{k_p\overline{k}_d}}
 = \Delta\mu=\mu_{ATP}-\mu_{ADP}-\mu_{P_i}
\end{equation}
}
where we have set $k_bT$ to unity and $\mu$ refers to the chemical potential.
Expressing the entropy production in terms of the steady state probability flux $J$ (see the proofs) we obtain
{ 
\begin{equation}
 \sigma =
 J\log{\frac{\overline{k}_pk_d}{k_p\overline{k}_d}}
\end{equation}
}
which shows that the average rate of entropy production corresponds 
to the rate of free energy consumption related to ATP hydrolysis (see e.g. \cite{hill}).
We recall that under physiological conditions the energy obtained by the hydrolysis of a molecule of ATP is of about $20 k_bT$ \cite{lang14,voet}. 
We observe numerically that if we consider several proteins the free energy consumption
scales linearly with the number of proteins $N$.

Let us now focus on the information functional (\ref{eq:info}). From the general results derived in the previous sections we know that 
it will start
from zero and, if the measurement begins when the system is at the steady state, increase monotonically in time. 
The initial growth can be obtained by a 
Taylor expansion of Eq.~(\ref{eq:learn}) which yields
\begin{equation}\label{eq:bc_learn}
 \lim_{\Delta t \to 0} \frac{\left\langle {\cal I}_1\right \rangle}{\Delta t} = \sum_{y,y} p(y) \chi(y \to y')
 D_{KL} \left( p(x'|y) \Bigr|\Bigr| p(x'|y') \right) +O(\Delta t)
\end{equation}
and, as shown in the proofs section, corresponds
to the learning rate discussed in \cite{horowitz_esposito,barato14,hartich14}. 
Such linear increase will gradually slow down
until a saturation level is reached.
The numerical evaluation of the acquired information  is shown in Fig.~\ref{fig:numerics_many}.
We find that it displays a dependence on the transition rates which differs from the one of the initial learning rate 
as it results from a balance between learning the more recent history of the receptor occupancy and
forgetting the older one.
 
To be more quantitative let us now focus on the case in which the sensory layer
is made of a single receptor and the memory device of several proteins.
We find that (see Fig.~\ref{fig:numerics_many}.a and Fig.~\ref{fig:numerics_many}.b)
when the number of proteins is
large, the information added by the memory device
scales as the Shannon entropy of the protein distribution (which itself scales as the logarithm of the number states the memory can take, i.e. $\log{(N+1)}$). 
The proportionality factor
(always smaller than $1$) depends on the specific rates.
This confirms the intuition  
that a larger number of proteins provides a larger memory and hence
allows to store more information about the history of the receptor.

\begin{figure}[!h]
\centering
\includegraphics[width=0.8\textwidth]{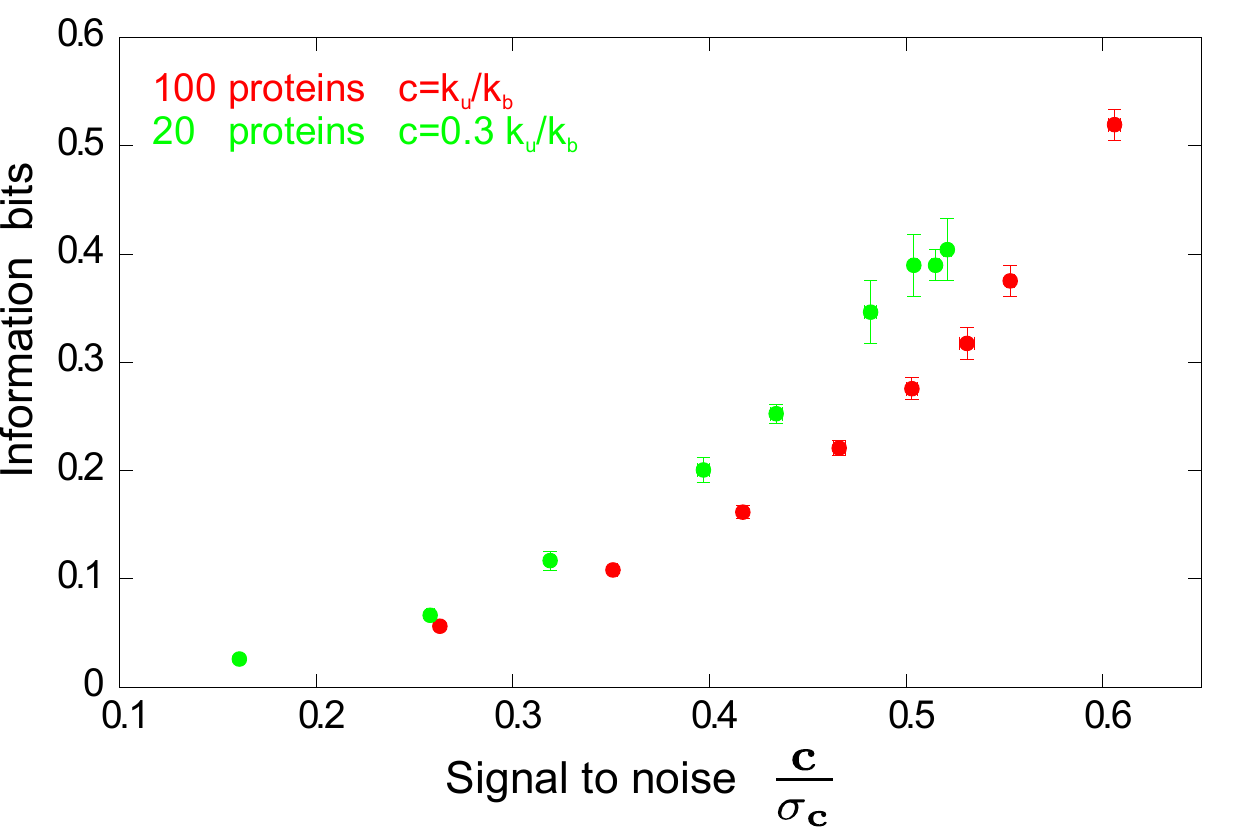}
\caption{Information acquired in the long time limit vs signal to noise. The green points refer to the case with $20$ proteins and a concentration of $c=0.3k_u/k_b$.
The red points are obtained for $100$ proteins and a concentration of $c=k_u/k_b$. All rates but $\overline{k}_p$ are kept equal $k_b=k_u=k_p=k_d=\overline{k}_d=0.2$. 
For the green points ($20$ proteins) $\overline{k}_p$ takes values 
$\overline{k}_p=1.5\,k_u\,,\,2\,k_u\,,\,2.5\,k_u\,,\,3.75\,k_u\,,\,5\,k_u\,,\,10\,k_u\,,\,20\,k_u\,,\,40\,k_u\,,\,80\,k_u$
and for the red points ($100$ proteins) it takes values 
$\overline{k}_p=1.25\,k_u\,,\,1.375\,k_u\,,\,1.5\,k_u\,,\,1.125\,k_u\,,\,1.75\,k_u\,,\,1.875\,k_u\,,\,2\,k_u\,,\,2.5\,k_u$. For the chosen parameters
set lower values of the
rate $\overline{k}_p$  imply lower values of the signal to noise.
Following Ref. \cite{mehta}
the signal to noise is evaluated as $\frac{c}{\sigma_c}=\frac{\partial \overline{n}}{\partial c}\frac{c}{\sigma_n}$ where
$\overline{n}$ is the average number of phosphorylated and $\sigma_n$ its numerically obtained standard deviation.}\label{fig:signoise}
\end{figure}

The information we are considering provides a measure of the cell ability to learn about
the environment. Such ability has often been evaluated in terms of the accuracy in determining the 
concentration of an external ligand (see e.g. Refs. \cite{berg_purcell,bialek05,endres09,mehta}).
In general these two measures are not in a one-to-one relation and subtleties may arise especially due 
to the finite size effects of the protein pool of our example. However, when the finite size effects are negligible
a higher information is associated to a higher signal to noise as shown in fig. \ref{fig:signoise}.

 \begin{figure}[!t h]
 \includegraphics[
width=0.8\columnwidth
]{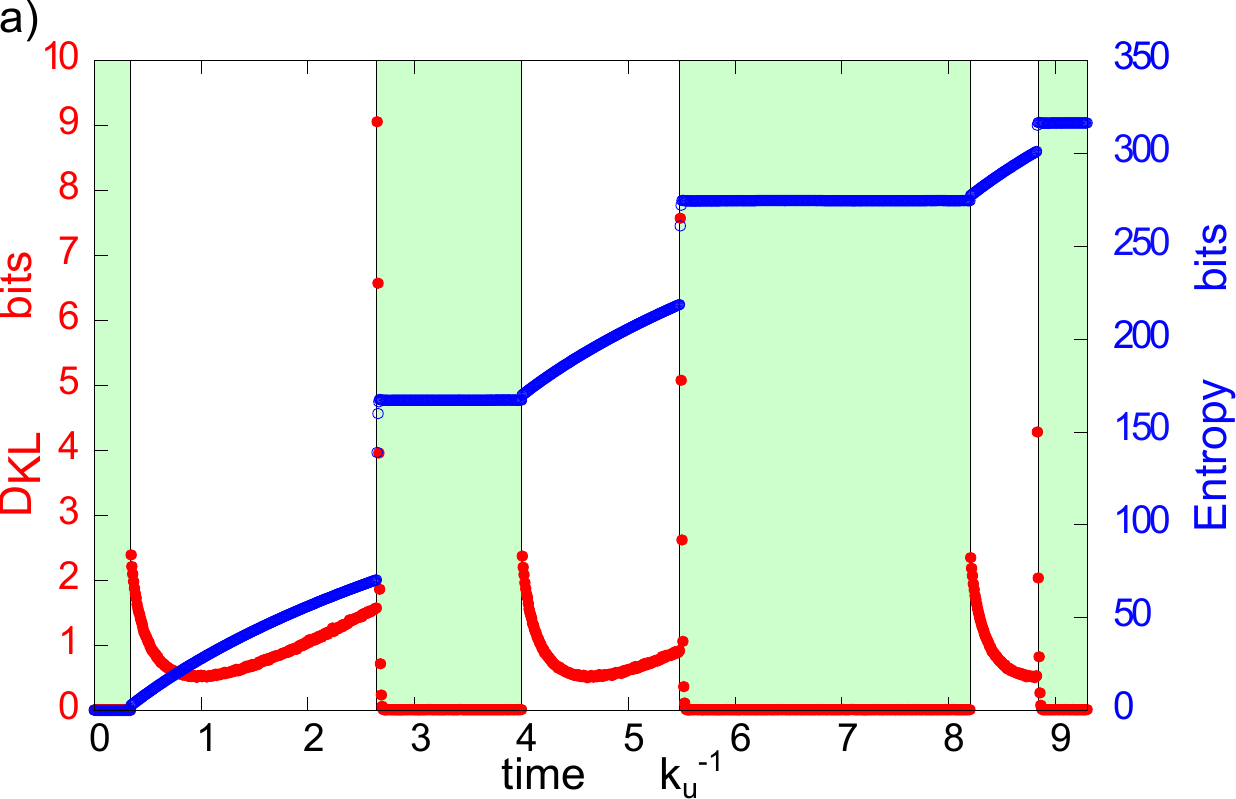}\\
\includegraphics[
width=0.8\columnwidth
]{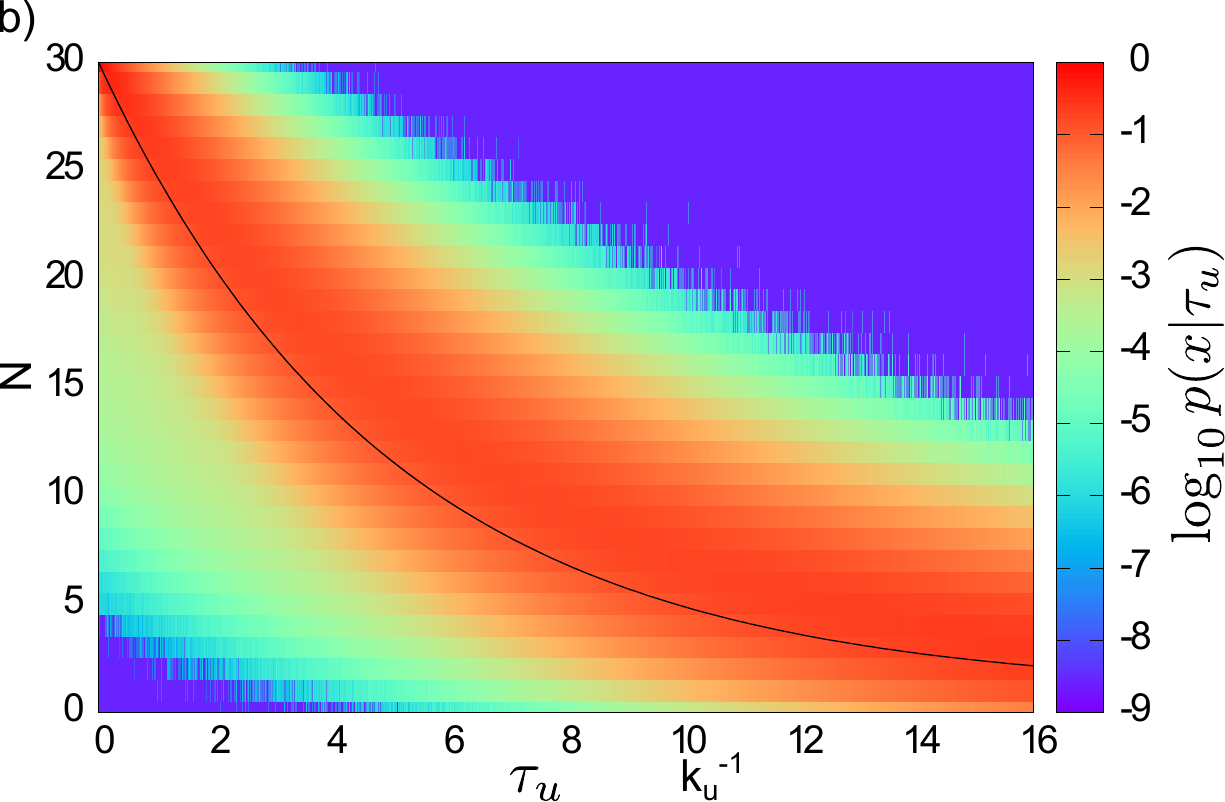}
\caption{Single receptor trajectory: the erasure procedure.
(a) We consider an example of a receptor trajectory consisting of a sequence of bound and unbound intervals.
The bound periods are shown by shading the background. We monitor the evolution of the proteins distribution following 
the receptor dynamics. By doing so we are able to compute how much entropy is produced on average by the proteins in the process of tracking
the receptor state. The blue line depicts such entropy production which is defined as
${\cal S}_t=-\langle \log p_{ss}(x_t|y_t) \rangle+\langle \log p_{ss}(x_0|y_0)\rangle+
\int_0^t \sigma_{t'} dt'$ where $ \sigma_t=\sum_{x,y,x',y'} p_{t}(x|y_{[0,t]})K(x,y \to x',y') \log \frac{K(x,y \to x',y')}{K(x',y' \to x,y)}$
is the single receptor trajectory analog of Eq.~(\ref{eq:bc_entro}) and the averages
are taken over $p(x_t|y_{[0,t]})$. For the sake of convenience we have set the Boltzmann constant to unity.
We also evaluate the amount of information harvested along this receptor trajectory and draw it in red. 
For a single trajectory this corresponds to the Kullback Leibler distance
between the protein distribution conditioned on the receptor trajectory and the one at the steady state conditioned on the
current receptor state:
$D_{KL}\left(p(x_t|y_{[0,t]})\big |\big|p_{ss}(x_t|y_t)\right)$ expressed in bits.
The parameter set corresponds to the erasure procedure discussed in the text and 
in Fig.~\ref{fig:numerics_many} with $c=k_u/k_b$ and $30$ proteins. The name erasure 
comes from the fact that, as shown, when the receptor is bound information is rapidly reset to zero.\\
 (b) We consider how the proteins probability when the receptor is unbound changes with  the the amount of time passed from the latest unbinding event $\tau_u$.
 As expected, for long unbound intervals, it approaches the equilibrium distribution set by the rates of the inactive receptors. 
We report a heat map of $\log_{10}{p(x|\tau_u)}$, the logarithm of the protein probability 
when the receptor is unbound conditioned on $\tau_u$  in units of $k_u$.
The solid line represents
the mean value of the distribution. The ligand concentration is set to $c=0.2k_u/k_b$.
}

\label{fig:numerics_single}

\end{figure}

 Let us now turn our attention to single trajectories of the receptor. We want to
determine how, given a specific sequence of binding and unbinding events, the proteins
follow its evolution and keep track of it.
We therefore obtain
 the probability distribution
of the proteins and monitor how it changes along the receptor transitions.
This allows us to compute how much information about the receptor history the memory is encoding
and how much entropy it is producing. The results for a typical
trajectory are shown in Fig.~\ref{fig:numerics_single}.a. 
From Eq.~(\ref{eq:info_kl}) we see that, along a single trajectory, information corresponds to the Kullback-Leibler
distance between the probability distribution of the proteins
conditioned on that trajectory and the steady state one 
conditioned only on the current receptor state.
Between receptor transitions the phosphorylation rates are constant and the probability of 
the memory states tends to the
equilibrium distribution set by these fixed rates. As a transition occurs the rates change and the distribution
seeks to reach the newly set equilibrium. This relaxation to different equilibria, driven
by the changes in the receptor state produces entropy. At the same time when a transition
takes place the current protein distribution 
may be very different 
from the one conditioned on the receptor state. Such difference encodes information on the history of the receptor.
This is the contribution added by the memory layer
to the information on the the current receptor state. 
In order to gain more insight let us consider 
a specific and illustrative case that we refer to as the erasure procedure.
We set the transition rates in a way such that when the receptor is bound
all the proteins in the pool are very rapidly phosphorylated, thereby erasing
the dependence on the previous receptor state. 
Fig.~\ref{fig:numerics_single}.a indeed shows that when the receptor is
bound information is rapidly reset to zero.
When the unbinding
occurs, the proteins are gradually dephosphorylated in time. 
We then expect to have most of the proteins phosphorylated when the 
receptor is bound and much fewer as the receptor sits in the unbound state 
(see Fig.~\ref{fig:numerics_single}.b). In this case the information on 
the history of the receptor consists in the knowledge about how long ago the last unbinding event took place,
given that the receptor is currently unbound.

 \paragraph{Conclusions and discussion.---}
 We have considered a stochastic measurement device that 
 stores information about the environment by monitoring the history
 of detections made by its sensory component. We 
 have derived an integral fluctuation theorem connecting the amount of entropy produced by the system
 with the information about the history of the sensory layer that is added by the memory device. 
 As a consequence of this result, we have shown that the average amount of information
 is bounded by the average thermodynamic entropy production by the process. A byproduct is that if the measurement
 device obeys detailed balance the device does not contain more information about the environment than its sensory
 component alone, making superfluous the addition of a memory component. When the external environment does not vary
 in time, a working measurement device eventually reaches a steady state and produces entropy linearly in time, on average. 
 However, the acquired information cannot grow beyond the limit set by the Shannon entropy of the memory component. This means
 that, for static environments, the system keeps dissipating
energy without possibly refining the measurement. Indeed, energy is continuously spent in the irreversible process of updating 
information on a finite memory, and the simple monitoring of the environment is expensive. 
In closing, we would like to point out that by allowing the memory device to exert a feedback
on the sensory layer opens the possibility of gathering information without introducing dissipation. 
However, this comes at the cost of perturbing the sensory device and entangles the information about the external environment with
the internal state of the device. 

 \section*{Proofs.}\label{sec:proofs}
 The  key observation for proving our results is that for a process in which the memory device $x$  does not exert any feedback on
 the sensory layer $y$, the $x$ sequence {\it conditioned} on the history of $y$ is a Markov process itself.
This follows from 
\begin{eqnarray}
\fl p(x_n|y_0,\ldots,y_n)=\sum_{x_0,\ldots,x_{n-1}} \frac{p(x_0,y_0; \ldots; x_n,y_n)}{p(y_0,\ldots,y_n)}\\\nonumber
\fl =\sum_{x_0,\ldots,x_{n-1}} \frac{p(x_{n},y_n|x_{n-1},y_{n-1}) p(x_0,y_0; \ldots; x_{n-1},y_{n-1})}{p(y_0,\ldots,y_n)} 
\\\nonumber\fl
= \sum_{x_0,\ldots,x_{n-1}}\frac{p(x_{n}|x_{n-1},y_n,y_{n-1})p(y_n|y_{n-1}) p(x_0,y_0; \ldots; x_{n-1},y_{n-1})}{p(y_0,\ldots,y_{n-1})p(y_n|y_{n-1})}
\\\nonumber\fl
= \sum_{x_{n-1}} p(x_{n}|x_{n-1},y_{n-1},y_n) p(x_{n-1}|y_0,\ldots,y_{n-1})
\end{eqnarray}
where we remark that it is the fact that $x$ does not influence the evolution of $y$ that allows to write
$p(x_{k+1},y_{k+1}|x_k,y_k)=p(x_{k+1}|x_k,y_{k+1},y_k)p(y_{k+1}|y_k)$.
\paragraph{Information as a Kullback-Leibler distance.---}
%
By definition one has
\begin{eqnarray}
 \fl \left\langle {\cal I}_n\right \rangle 
 &= \sum_{x_{0},y_{0}\ldots x_{n}, y_{n}} p(x_{0},y_{0}\ldots x_{n}, y_{n})  \log \frac{p(x_n|y_0,\ldots,y_{n})}{p(x_n|y_n)} \\\nonumber
 \fl&= \sum_{y_{0}\ldots  y_{n}} p(y_{0}\ldots  y_{n})\sum_{x_{0}\ldots x_{n}} p(x_{0}\ldots x_{n}|y_{0}\ldots, y_{n}) 
 \log \frac{p(x_n|y_0,\ldots,y_{n})}{p(x_n|y_n)} 
 \\ \nonumber
\fl &=\sum_{y_{0}\ldots  y_{n}} p(y_{0}\ldots  y_{n})\sum_{x_{n}} p(x_{n}|y_{0}\ldots, y_{n}) 
 \log \frac{p(x_n|y_0,\ldots,y_{n})}{p(x_n|y_n)}
  \\ \nonumber
 \fl&= \sum_{y_{0}\ldots y_{n}} p(y_0,\ldots,y_{n}) D_{KL} \left( p(x_n|y_0,\ldots,y_{n}) \Bigr|\Bigr| p(x_n|y_n) \right)\;.
\end{eqnarray}
\paragraph{Jarzynski identities.---}
We show here the derivation of {\it quenched} Jarzynski identities, i.e. averages  taken at a fixed sequence of states $y_0,\ldots,y_n$. The annealed results follow directly by averaging over the $y$ process.
For the entropy production we have that
\begin{eqnarray}
&\fl\left\langle e^{-{\cal S} +{\cal S}^{Y}} \bigr| y_0,\ldots,y_n \right\rangle  
\\ \nonumber
&\fl=\sum_{x}  \frac{p(x_0,y_0,\ldots,x_n,y_n)}{p(y_0,\ldots,y_n)}
\underbrace{\frac{p(x_n,y_n)}{p(x_0,y_0)}
\prod_{k=0}^{n-1} \frac{p(x_{k}|x_{k+1},y_{k+1},y_{k})p(y_{k}| y_{k+1})}{ p(x_{k+1}|x_k,y_k,y_{k+1})p(y_{k+1}|y_k) }}_{e^{-\cal S}} 
\underbrace{ \frac{p(y_0;\ldots;y_n)}{p_B(y_n;\ldots;y_0)}}_{e^{\cal {\cal S}^{Y}}} 
\\\nonumber
&\fl= \sum_{x} \left( p(x_n|y_n) \prod_{k=0}^{n-1} p(x_{k}|x_{k+1},y_{k+1},y_{k}) \right) \left(p(y_n) \prod_{k=0}^{n-1}  p(y_{k}|y_{k+1}) \right)
\frac{1}{p_B(y_n;\ldots;y_0)} 
\\ \nonumber
&\fl= \frac{p_B(y_n;\ldots;y_0)}{p_B(y_n,\ldots,y_0)}=1
\end{eqnarray}
where $p_B(y_n,\ldots,y_0)$ denotes the probability of observing the sequence of states
$y_n,\ldots,y_0$ i.e. the backward evolution.
The fluctuation theorem expressed in Eq.~(\ref{eq:IFT}) is obtained by
\begin{eqnarray}
\fl\left\langle e^{-{\cal S} +{\cal S}^{Y}+ {\cal I}}\bigr| y_0,\ldots,y_n \right\rangle 
\\\nonumber
\fl=\sum_{x}  \frac{p(x_0,y_0,\ldots,x_n,y_n)}{p(y_0,\ldots,y_n)}
\underbrace{\frac{p(x_n,y_n)}{p(x_0,y_0)}
\prod_{k=0}^{n-1} \frac{p(x_{k}|x_{k+1},y_{k+1},y_{k})p(y_{k}| y_{k+1})}{ p(x_{k+1}|x_k,y_k,y_{k+1})p(y_{k+1}|y_k) }}_{e^{-\cal S}} \times
\\\nonumber
\fl\times
\underbrace{ \frac{p(y_0;\ldots;y_n)}{p_B(y_n;\ldots;y_0)}}_{e^{\cal {\cal S}^{Y}}} 
\underbrace{\frac{p(x_n|y_0,\ldots,y_n)}{p(x_n|y_n)}}_{e^{\cal I}}
\\ \nonumber
\fl= \sum_{x} \left( p(x_n|y_0,\ldots,y_n) \prod_{k=0}^{n-1} p(x_{k}|x_{k+1},y_{k+1},y_{k}) \right) \left(p(y_n) 
\prod_{k=0}^{n-1}  p(y_{k}|y_{k+1}) \right)\frac{1}{p_B(y_n;\ldots;y_0)}
\\\nonumber
\fl
= \frac{p_B(y_n;\ldots;y_0)}{p_B(y_n,\ldots,y_0)}=1
\end{eqnarray}
Finally, the fluctuation relation for information which, exploiting Jensen inequality,
allows to write Eq.~(\ref{eq:ineq}) follows from
\begin{eqnarray}
\fl
\left\langle e^{- {\cal I}} \bigr| y_0,\ldots,y_n \right\rangle &= \sum_{x} \frac{p(x_0,y_0,\ldots,x_n,y_n)}{p(y_0,\ldots,y_n)}
\underbrace{\frac{p(x_n|y_n)}{p(x_n|y_0,\ldots,y_n)}}_{e^{-\cal I}}
\\\nonumber
&= \sum_{x_n}  p(x_n|y_0,\ldots,y_n)  \frac{p(x_n|y_n)}{p(x_n|y_0,\ldots,y_n)} = 1\;.
\end{eqnarray}

\paragraph{Information at the steady state.---}

At the steady state it is possible to show that the information defined in Eq.~(\ref{eq:info}) is an 
increasing function of time.
Indeed, by the data processing inequality (cf Ref. \cite{cover}) one  has
that 
$I(x_n:\{y_0,\ldots,y_n\})\ge I(x_n:\{y_k,\ldots,y_n\})$ 
for all $k=0,\ldots, N$. For $k=N$ this implies positivity $\langle {\cal I}_n \rangle \ge 0$. Subtracting $I(x_n:y_n)$ 
from both sides of the data-processing inequality yields $\langle {\cal I}_n \rangle \ge I(x_n:\{y_k,\ldots,y_n\})-
I(x_n:y_n)$. At the steady state, $I(x_n:\{y_k,\ldots,y_n\})=I(x_{n-k}:\{y_0,\ldots,y_{n-k}\})$ and $I(x_n:y_n)=I(x_{n-k}:y_{n-k})$.
This leads to $\langle {\cal I}_n \rangle \ge \langle {\cal I}_{n-k} \rangle$ which for $k=1$ shows that at the steady state the
information is an increasing function of time. 
The growth of information is not indefinite, though, since the information is bounded from above by 
$\langle {\cal I}_n \rangle \le I(x_n:\{y_0,\ldots,y_n\}) \le H(x_n)$.
\paragraph{The equilibrium case.---}
Since the evolution of the sensory layer $y$ is independent of the state of the memory device $x$
we expect the current state of the memory to have no information on the future evolution of
the sensory layer: $i(x_0:\{y_0,...,y_n\})=i(x_0:y_0)$.
Indeed
\begin{eqnarray}
\fl \frac{p(x_0|y_0,...,y_n)}{p(x_0|y_0)}&= \sum_{x_1,...,x_n}\frac{p(x_0,y_0;...;x_n,y_n)}{p(x_0|y_0)p(y_0,...,y_n)}
\\\nonumber
&=\sum_{x_1,...,x_n}\prod_{k=0}^{n-1}\frac{p(x_{k+1}|x_k,y_k,y_{k+1})p(y_{k+1}|y_k)}{p(y_{k+1}|y_k)}
\nonumber \\
&=\sum_{x_1,...,x_n}\prod_{k=0}^{n-1}p(x_{k+1}|x_k,y_k,y_{k+1})=1 \;. \nonumber
\end{eqnarray}
More generally, for any $k\leq n$, $p(x_k|y_k,...,y_n)=p(x_k|y_k)$.
Notice that this does not imply that $x$ is independent of the future history of $y$ that would be $p(x_k|y_{k+1},...,y_n)=p(x_k)$.\\
\\
If the full system is at equilibrium the condition of detailed balance holds
\begin{equation}
\frac{p(x_{k+1},y_{k+1}|x_k,y_k)}{p(x_k,y_k|x_{k+1},y_{k+1})}=\frac{p_{eq}(x_{k+1},y_{k+1})}{p_{eq}(x_k,y_k)}
\end{equation}
and we can show that
\begin{eqnarray}
\fl p_{eq}(x_n|y_0,...,y_n)&=\frac{p_{eq}(x_n,y_0,...,y_n)}{p_{eq}(y_0,...,y_n)}\\
&=
\sum_{x_0,...,x_{n-1}}\frac{p_{eq}(x_0,y_0)}{p_{eq}(y_0)}\prod_{k=0}^{n-1}\frac{p(x_{k+1},y_{k+1}|x_k,y_k)}{p(y_{k+1}|y_k)} \nonumber\\ 
&= \sum_{x_0,...,x_{n-1}}\frac{p_{eq}(x_0,y_0)}{p_{eq}(y_0)}\prod_{k=0}^{n-1}\frac{p(x_k,y_k|x_{k+1},y_{k+1})\frac{p_{eq}(x_{k+1},y_{k+1})}{p_{eq}(x_k,y_k)}}
{p(y_{k}|y_{k+1})
\frac{p_{eq}(y_{k+1})}{p_{eq}(y_{k})}
} \nonumber\\ 
&= \frac{p_{eq}(x_n,y_n)}{p_{eq}(y_n)}\sum_{x_0,...,x_{n-1}}\prod_{k=0}^{n-1}\frac{p(x_k,y_k|x_{k+1},y_{k+1})}
{p(y_{k}|y_{k+1})
}=p_{eq}(x_n|y_n)\;.\nonumber
\end{eqnarray}

For bipartite systems which do not display transitions that change
simultaneously the states of both variables we prove the stronger relation: $i(x_n:y_n)=0$.

In the absence of feedback by detailed balance one has
\begin{eqnarray}
\fl p_{eq}(x',y')&=&p_{eq}(x',y)\frac{p(x',y'|x,y)}{p(x',y|x',y')}= 
p_{eq}(x',y)\frac{p(x'|x',y,y')p(y'|y)}{p(x'|x',y',y)p(y|y')} \nonumber \\
\nonumber \\
&=&p_{eq}(x',y)\frac{p(x'|x',y,y')p_{eq}(y')}{p(x'|x',y',y)p_{eq}(y)}\;.
\end{eqnarray}
For bipartite chains, if $y'\neq y$ than $x'=x$ so that $p(x|x,y,y')=1$
and
\begin{equation}
p_{eq}(x'|y)=p_{eq}(x'|y')=p_{eq}(x')
\end{equation}
showing that at equilibrium for a bipartite system with no feedback 
the memory device and the sensory layer are independent.
\paragraph{Learning rate.---}
Combining the definitions in Eq.~(\ref{eq:info_kl}) and Eq.~(\ref{eq:learn}) one has
\begin{eqnarray}
\left\langle {\cal I}_1\right \rangle 
= \sum_{y,y'}  p(y;y')  D_{KL} \left( p(x'|y,y') \Bigr|\Bigr| p(x'|y') \right)\;.
\end{eqnarray}
In the continuous-time limit
\[
p(x'|x,y,y') = \delta_{xx'}(1-\rho(x|y)\Delta t) + \xi(x\to x'|y) \Delta t + o(\Delta t)
\]
where 
\[
\rho(x|y)=\sum_{x'} \xi(x\to x'|y) 
\]
and $\xi(x\to x)=0$.
Similarly
\[
p(y'|y) = \delta_{yy'} (1-\omega(y)\Delta t) + \chi(y \to y') \Delta t + o(\Delta t)
\]
Therefore
\[
p(x'|y,y') = \sum_x \left[ \delta_{xx'}(1-\rho(x|y)\Delta t) + \xi(x\to x'|y) \Delta t \right] p(x|y)\]
\[
=p(x'|y) + \left [\sum_x  \xi(x\to x'|y) p(x|y) - \rho(x'|y) p(x'|y) \right] \Delta t
\]
and
\begin{eqnarray*}
\fl D_{KL} \left( p(x'|y,y') \Bigr|\Bigr| p(x'|y') \right) 
\\
\fl=\sum_{x'}  \left[ p(x'|y) + \left (\sum_x  \xi(x\to x'|y) p(x|y) - \rho(x'|y) p(x'|y) \right) \Delta t \right]\times
\\
\fl\times\log \frac{p(x'|y) + \left [\sum_x  \xi(x\to x'|y) p(x|y) - \rho(x'|y) p(x'|y) \right] \Delta t}{p(x'|y')}
\\
\fl= \sum_{x'}  \left[ p(x'|y) + \left (\sum_x  \xi(x\to x'|y) p(x|y) - \rho(x'|y) p(x'|y) \right) \Delta t \right] \times
\\
\fl\times
\log \left[ \frac{p(x'|y)}{p(x'|y')}\left(1 + \frac{ \left [\sum_x  \xi(x\to x'|y) p(x|y) - \rho(x'|y) p(x'|y) \right] \Delta t}{p(x'|y)} \right)\right]
\\
\fl= D_{KL} \left( p(x'|y) \Bigr|\Bigr| p(x'|y') \right) 
+ \Delta t \sum_{x'} \left( 1+ \log{\frac{p(x'|y)}{p(x'|y')}}\right) \left (\sum_x  \xi(x\to x'|y) p(x|y) - \rho(x'|y) p(x'|y) \right)
\\
 \fl=D_{KL} \left( p(x'|y) \Bigr|\Bigr| p(x'|y') \right) 
+ \Delta t \sum_{x'} \log{\frac{p(x'|y)}{p(x'|y')}} \left (\sum_x  \xi(x\to x'|y) p(x|y) - \rho(x'|y) p(x'|y) \right)
\end{eqnarray*}
Plugging this result into the expression for the learning rate
\[\fl
\left\langle {\cal I}_1\right \rangle = \sum_{y,y'} p(y) \left[ \delta_{yy'} (1-\omega(y)\Delta t) + \chi(y \to y')
\Delta t \right]\times
\]
\[\fl
\left[D_{KL} \left( p(x'|y) \Bigr|\Bigr| p(x'|y') \right) + \Delta t \sum_{x'} \log{\frac{p(x'|y)}{p(x'|y')}} \left (\sum_x  \xi(x\to x'|y) p(x|y) - \rho(x'|y) p(x'|y) \right)\right]
\]
noticing that for $y=y'$ the second factor vanishes, 
one obtains
\begin{equation}
 \lim_{\Delta t \to 0} \frac{\left\langle {\cal I}_1\right \rangle}{\Delta t} = \sum_{y,y} p(y) \chi(y \to y')
 D_{KL} \left( p(x'|y) \Bigr|\Bigr| p(x'|y') \right) +O(\Delta t)
\end{equation}
which is Eq.~(\ref{eq:bc_learn}). Finally, exploiting the fact that at the steady state
by definition we have that:
\[
 \fl\sum_{x'} \xi(x'\to x|y)p_{ss}(x',y)- \xi(x\to x'|y)p_{ss}(x,y)=
 \sum_{y'} -\chi(y'\to y)p_{ss}(x,y')+ \chi(y\to y')p_{ss}(x,y)
\]
we can rewrite Eq.~(\ref{eq:bc_learn}) as
\begin{eqnarray}\label{eq:info_rate_deriv}
\fl  \lim_{\Delta t \to 0} \frac{\left\langle {\cal I}_1\right \rangle}{\Delta t}=
  \sum_y p_{ss}(y)\sum_{x\,x'}\left(\xi(x'\to x|y)p_{ss}(x'|y)\right)\log{\frac{p_{ss}(x|y)}{p_{ss}(x'|y)}}+O(\Delta t)
\end{eqnarray}
which is the expression derived in \cite{barato14}.

\paragraph{Entropy production and chemical forces.---}
Let us focus on the simple case of one receptor and one protein ($R=1$, $N=1$).
In analogy with Ref.~\cite{barato14} we consider  the
phosphorylation involving the active receptor to be associated with ATP consumption and the one with the inactive
receptor to require no ATP. We remark that, a complete description should account for 
the different chemical reactions connecting the states as done by the authors of Ref.~\cite{barato13pre}.
This results in the thermodynamic constraint 
{ 
\begin{equation}
 \log{\frac{\overline{k}_pk_d}{k_p\overline{k}_d}}
 = \Delta\mu=\mu_{ATP}-\mu_{ADP}-\mu_{P_i}
\end{equation}
}
where we have set $k_bT$ to unity.
From equation (\ref{eq:bc_entro}) we can write the average entropy production rate as
\begin{eqnarray}
 \sigma &=&\sum_{i,j} p_i K(i\to j)\log\frac{ K(i\to{j})}{K(j\to{i})}\\\nonumber
 &=&\frac{1}{2}\sum_{i,j}\left(p_i K(i\to{j})-p_j K(j\to i)\right)\log\frac{ K(i\to{j})}{K(j\to{i})}\,.
\end{eqnarray}
The flux between state $i$ and $j$ is defined as:
\begin{equation}
 J(i\to j)=p_i K(i\to j)-p_j K(j\to i)\;.
\end{equation}
At the steady state the flux between each pair of states are constant and for the 4 states network we are considering the fluxes
in the clockwise direction are all equal:
\begin{eqnarray}
 \fl&&J(x=0,y=0\to x=0,y=1)=J(x=0,y=1\to x=1,y=1)\\\nonumber
 \fl&=&J(x=1,y=1\to x=1,y=0)=J(x=1,y=0\to x=0,y=0)=J\,.
\end{eqnarray}
The entropy production then reads:
{ 
\begin{equation}
 \sigma =
 J\log{\frac{\overline{k}_pk_d}{k_p\overline{k}_d}}=
 J\Delta \mu
\end{equation}}
coinciding with the ATP consumption. Explicitly, we have that the expression for the flux reads
{ 
 \begin{equation}
 J=
 \frac{ c\,k_b  k_u \left(\overline{k}_p k_d -  k_p \overline{k}_d \right)}
 {\left( c\,k_b  +k_u \right)\left[c\,k_b  \left(\overline{k}_d +\overline{k}_p \right)+k_u \left(k_d + k_p \right)+
 \left(\overline{k}_d +\overline{k}_p \right)\left(k_d + k_p \right) \right]}\,.
\end{equation}
}

\paragraph{Acknowledgements}
SB acknowledges ICTP for hospitality.

\end{document}